\documentclass[A4paper,10pt]{article}
\usepackage{amsfonts}
\usepackage[english]{babel}

\begin{document}
\title{\bf Noncommutative fermions and Morita equivalence}
\author{D.H.~Correa \thanks{CONICET} \, and \,
E.F.~Moreno \thanks{Associated with CONICET}
\\
{\normalsize\it Departamento de F\'\i sica, Universidad Nacional
de La Plata}\\ {\normalsize\it C.C. 67, 1900 La Plata, Argentina}}

\maketitle

\begin{abstract}
We study the Morita equivalence for fermion theories on
noncommutative two-tori. For rational values of the $\theta$
parameter (in appropriate units) we show the equivalence between
an abelian noncommutative fermion theory and a nonabelian theory
of twisted fermions on ordinary space. We study the chiral anomaly
and compute the determinant of the Dirac operator in the dual
theories showing that the Morita equivalence also holds at this
level.
\end{abstract}

%



\section{Introduction}
Noncommutative Field Theories (NCFT) have attracted much attention in the
last years because they naturally arise as some low energy limit of open
string theory and as the compactification of M-theory on the torus \cite{CDS},\cite{SW}.

A significant feature of the noncommutative field theory is the
Morita duality between noncommutative tori. This duality is a
powerful mathematical result that establishes a relation, via an
isomorphism, between two noncommutative algebras. Of particular
importance are the algebras defined on the noncommutative torus,
where it can be shown that are Morita equivalent if the
corresponding sizes of the tori and the noncommutative parameters
are related in a specific way.

There has been several results in the literature about the Morita
equivalence of NCFT but principally focused on noncommutative
gauge theories and describing mostly classical or semiclassical
aspects of them \cite{schwarz}-\cite{GSV}. In particular there has
been relatively very little work on other than gauge theories or
in the quantum aspects of the equivalence. Central questions as if
there are ``Morita anomalies" are still open.

In this work we want to fill some gaps on the subject. First we are going to
establish the Morita equivalence for fermion theories. We are going to show
that there is a well defined isomorphism between the correlation functions of
fermions on a noncommutative torus and those of a non-abelian fermion theory
on ordinary space.  Also we are going to analize the effect of the Morita map
on the chiral anomaly and compute and compare the fermionic determinant of
dual theories. Finally we are going to discuss the bosonization of fermion
theories defined on dual noncommutative tori.

\section{Morita equivalence for fermionic fields}

The Morita equivalence is an isomorphism between noncommutative algebras that
conserves all the modules and their associated structures. Let us consider
the noncommutative torus $T^2_{\theta}$ and for simplicity, of radii $R$. The
coordinates satisfy the commutation rule
\begin{equation} [x_1,x_2]=i \; \theta \end{equation} %
An associative algebra of smooth functions over $T^2_{\theta}$ can
be realized through the Moyal product
\begin{equation} f(x)* g(x) =\left.  \exp\left( \frac{i\theta}{2} (\partial_{x_1}
\partial_{y_2} - \partial_{x_2} \partial_{y_1}) \right)\;
f(x) \; g(y) \right|_{y=x} \end{equation} %

It is convenient to decompose the elements of the algebra in their
Fourier components. However, when dealing with fermions defined on
a torus we must be aware that they can have different spin
structures associated to any of the compact directions. For the
torus we can have four different spin structures characterized as
follows
\begin{equation}
\psi_(x_1+R,x_2) = e^{2 \pi i \alpha_1}\;\psi(x_1,x_2)\; ,
\;\;\;\;\;\;
\psi_(x_1,x_2+R) = e^{2 \pi i \alpha_2}\; \psi(x_1,x_2)
\label{spinst}
\end{equation}
where $\alpha_1$ and $\alpha_2$ can take the values $0,1/2$. We
will call a fermion with boundary conditions (\ref{spinst}) as of
type $\vec \alpha=(\alpha_1,\alpha_2)\,$ and denote it $\psi_{\vec
\alpha}$.
Fermions with $\alpha_i=0 \; (i=1,2)$  are called Ramond
(R) and with $\alpha_i=1/2$ are called Neveu-Schwarz (NS).

The Fourier expansion of a fermion field of type $\vec \alpha$ has
the following form:
\begin{equation}\label{modos}
\psi_{\vec \alpha}=\sum_{\vec k}\psi^{\vec{k}}
U_{\vec{k}+\vec{\alpha}}\; ,\;\;\;\;{\rm with}\;\;\;\;
U_{\vec{k}} \equiv \exp \left(2\pi i\vec{k}\cdot\vec{x}/R\right)
\end{equation}
%
%
%

The Moyal commutator of the generators can be easily computed to give
\begin{equation}
[U_{\vec{k}+\vec{\alpha}},U_{\vec{k}'+\vec{\alpha}'}]=
-2i\sin\left(\frac{2\pi^2\theta}{R^2}(\vec{k}+\vec{\alpha})
\wedge(\vec{k}'+\vec{\alpha}')\right)
U_{\vec{k}+\vec{k}'+\vec{\alpha}+\vec{\alpha}'} \label{alg}
\end{equation}
where $\vec p \wedge \vec q = \varepsilon_{ij}\; p_i q_j$.

When the noncommutative parameter $\theta$ takes a the value
\begin{equation}
\theta =  \frac{4 M}{N} \frac{R^2}{2 \pi}
\end{equation}
being $M$ and $N$ relatively prime integers, an interesting
feature of the algebra generated by the $U_{\vec{k}+\vec{\alpha}}$
emerges. First, the infinite-dimensional algebra breaks up into
equivalence classes of finite dimensional subspaces. Indeed,
noticing that the elements $U_{N\vec{k}/2}$
generates the center of the algebra, we can decompose the momenta
in the form
\begin{equation}
2(\vec k' + \vec\alpha) = N\vec k + \vec n \;, \;\;\;\;\;0\leq n_1, n_2
\leq N-1
\end{equation}
and the whole algebra splits into equivalence classes classified
by the all possible values of $N \vec k$. Each class is itself a
subalgebra generated by the $N^2$ functions
$U_{\vec{n}+\vec{\alpha}}$
%
satisfying
\begin{equation}
[U_{\vec{n}+\vec{\alpha}},U_{\vec{n}'+\vec{\alpha}'}]=
-2i\sin\left(\pi\frac{M}{N}(2\vec{k} + 2\vec{\alpha})
\wedge(2\vec{k}' + \vec{2\alpha}')\right)U_{\vec{n} +
\vec{n}'+\vec{\alpha} + \vec{\alpha}'} \label{algebra}
\end{equation}
It is easy to show that the algebra (\ref{algebra}) is isomorphic to the
(complexification of the) Lie algebra $u(N)$. A N-dimensional representation
of this algebra can be constructed with ``shift" and ``clock" matrices
\cite{saraikin},\cite{gura},\cite{thooft},\cite{zachos}
\begin{equation} Q=\left(\begin{array}{ c c c c}
         1 & & &  \\  & \omega & &  \\
         & & . &  \\
          & & & \omega^{N-1} \\
        \end{array}\right)\;, \;\;\;\;\;\;
P=\left(\begin{array}{c c c c}
         0 & 1& & 0 \\
         & . & .& \\   & & . &1 \\
         1 & & &  0 \\
        \end{array}\right)
\label{pq}
\end{equation} %
where $\omega = \exp\left(\frac{2\pi iM}{N}\right)$. Indeed, the
matrices $ J_{\vec{n}} = \omega^{\frac{n_1 n_2}{2}} Q^{n_1} P^{n_2}$,
with $n_1,\,n_2=0,\cdots,N-1$, generates an algebra isomorphic to (\ref{algebra})
\begin{equation}
 [J_{\vec{n}},J_{\vec{m}}]=
 - 2i \sin\left(\pi \frac{M}{N}\vec{n}
 \wedge\vec{m}\right)J_{\vec{n}+\vec{m}}
\end{equation}
Thus, we have a map (Morita mapping) between the Fourier modes
defined on a noncommutative torus and functions taking values on
$u(N)$ defined on a commutative space:
\begin{equation}
\exp\left(2\pi i(\vec{k}+\vec{\alpha})\cdot
\hat{\vec{x}}/R)\right)\, \leftrightarrow \,\exp\left(2\pi
i(\vec{k}+\vec{\alpha})\cdot
\vec{x}/R\right)J_{2(\vec{k}+\vec{\alpha})} \label{morita}
\end{equation}
This mapping generates a mapping between fermion fields in the
following way. For the sake of simplicity, let us consider the
case $N=2N'$ in which the decomposing of the momenta is
$\vec k = N'\vec q + \vec n$  with $0\leq n_1,\,n_2 \leq N'$.
Then, we write the fermion field on the noncommutative torus
$T^{2}_{\theta}$ with spin structure $\vec\alpha$ in the form
\begin{equation}
\psi_{\vec \alpha}= \sum_{\vec q} \; \exp\left(2\pi i
N'\vec{q}\cdot\vec{x}/R\right)\; \sum_{\vec n =0}^{N'-1}
\psi^{\vec{q},\vec{n}} U_{\vec n+\alpha} \label{mapf}
\end{equation}
Now, using (\ref{morita}) is immediate to see that the Morita
correspondence between fermion fields is given by
\begin{equation}
\psi_{\vec\alpha} \;\; \leftrightarrow \;\;
\psi=\sum_{\vec{n}=0}^{N'-1}\chi^{(\vec{n})}J_{2\vec{n}+2\vec{\alpha}} \label{map1}
\end{equation}
and we have defined
\begin{equation}
\chi^{(\vec{n})} = \exp\left(2\pi i(\vec{n}+\vec{\alpha})\cdot
\vec{x}/R\right) \sum_{\vec{q}} \psi^{\vec{q},\vec{n}}\exp(2\pi i
N'\vec{q}\cdot\vec{x}/R) \label{map3}
\end{equation}
Notice that the fermion $\psi$ is defined in the dual torus of
size $R'=R/N'$ satisfying the boundary conditions
\begin{equation}
\psi(x_1+R',x_2)=\Omega_1^\dagger \cdot \psi(x_1,x_2) \cdot
\Omega_1\; , \;\;\;\;
\psi(x_1,x_2+R')=\Omega_2^\dagger \cdot \psi(x_1,x_2) \cdot \Omega_2
\label{bc}
\end{equation}
with
\begin{equation}\label{omega}
\Omega_1 = P^b,\;\;\; \Omega_2 = Q^{1/M}
\end{equation}
where $b$ is an integer satisfying $aN-bM =1$. While the field
components $\chi^{(\vec{n})}$ obey twisted boundary conditions
\begin{eqnarray}
\chi^{(\vec n)}(x_1 + R', x_2) &=& e^{2\pi i
(n_1 + {\alpha}_1)/N'} \chi^{(\vec n)}(x_1, x_2) \nonumber \\
\chi^{(\vec n)}(x_1, x_2 + R')&=& e^{2\pi i (n_2 +
{\alpha}_2)/N'}\chi^{(\vec n)}(x_1, x_2)
\end{eqnarray}
That is, the spin fields $\chi^{(\vec n)}$ have (twisted) spin structures
\begin{equation}
\left(\frac{n_1+\alpha_1}{N'}, \frac{n_2+\alpha_2}{N'}\right) \;,
\;\;\; \; n_1=0,\cdots,N'-1\;,\;\; n_2=0,\cdots,N'-1
\label{spinsts}
\end{equation}

Let us remark that integrating over the noncommutative torus $T^2_\theta$ is
equivalent to taking trace in the group U(N) and integrating over the dual
torus ${T^2}'$ simultaneously
\begin{equation}\label{integral}
\int_{T^2_\theta}\hat{\Gamma}(\vec x)\, d^2x =\frac{N'}{2} {\rm tr}_G
\int_{{T^2}'} \Gamma(\vec x)\, d^2x
\end{equation}

\section{Free fermions fields on the torus}

Consider, as warm up, a theory of a free Dirac fermion on the noncommutative
torus $T^2_{\theta}$. The action is given by
\begin{equation}
S=-\frac{1}{8\pi}\int_{T^2_\theta}\,  d^2x\, \bar{\psi}\ast ( \;
\not\!\partial + m) \psi \label{action1}
\end{equation}
%
%
For the case $2\pi\theta/R^2=4M/2N'$, we can use the Morita map
(\ref{morita}),(\ref{integral}) and we get
\begin{equation}
S = -\frac{N'^2}{8\pi}\sum_{\vec{n}=0}^{N'-1}\int_{{T^2}'} d^2x\,
\bar{\chi}^{(\vec{n})}( \; \not\!\partial + m)\chi^{(\vec{n})}
\label{action2}
\end{equation}
where $T'^2$ is the commutative torus of radii $(R/N',R/N')$. Thus, the
Morita equivalence establishes the relation between a theory of fermions with
spin structure $\vec \alpha$ defined on the noncommutative torus
$T^2_{\theta}$ and a theory of $N'^2$ fermions with spins structures (\ref{spinsts}),
defined on a commutative torus $T'^{2}$. Notice that the action
(\ref{action1}) is quadratic in the fields, and consequently independent of
$\theta$. Thus the equivalence between actions (\ref{action1}) and
(\ref{action2}) is {\sl independent} of $N'$.

At this point we have worked at classical level. We will see that the
equivalence also works at quantum level.
First we compute the partition function of both theories. As we stated above,
the action (\ref{action1}) is independent of $\theta$, so the $*$ product can
be replaced by the ordinary product. Hence, the partition function is the one
of a free Dirac fermion on an ordinary torus $T^2$, {\it i.e.\,}
\begin{equation}
Z={\rm det}_{T^2} \left( \not\!\partial + m \right)
\end{equation}
This determinant can be computed exactly for the case $m=0$, where
the theory becomes conformally invariant, and is given by
\cite{AGMV}, \cite{NS} \footnote{Strictly speaking this result is
valid only for $\vec \alpha \neq (0,0)$. If $ \alpha = (0,0)$ the
Dirac operator has a zero mode that has to be eliminated. For
simplicity from now on we will only consider the case $\vec \alpha
\neq (0,0)$.}
\begin{equation}
Z= \frac{\left|\vartheta\left[^{\alpha_1 -1/2}_{\alpha_2
-1/2}\right] (0,\tau)\right|^2}{\left|\eta(\tau)\right|^2}
\label{part1}
\end{equation}
where
\begin{equation}
\vartheta\left[^{a}_b\right](z,\tau)=\sum_{n\in {\mathbb Z}} \; e^{i\pi
\tau (n+a)^2 + 2\pi i (n+a)(z+b)^2} \; , \;\;\;\;
\eta(\tau) = e^{i{\frac{\pi \tau}{12}}} \prod_{n=1}^{\infty} \; (1-e^{2 \pi
i n \tau}) \nonumber
\end{equation}
are the Jacobi theta function and the Dedekind eta function respectively, and
$\tau = i$, the imaginary unity.

Now we compute the partition function of the theory defined by
(\ref{action2}). This theory corresponds to $N'^2$ free fermions with the
spin structures given in equation (\ref{spinsts}). The partition function is
given by
\begin{equation}
Z'= \prod_{n_1,n_2=0}^{N'-1}\left|\frac{\vartheta
\left[^{(\alpha_1 + n_1)/N'-1/2}_{(\alpha_2 + n_2)/N'-1/2}\right]
(0,\tau)}{\eta(\tau)}\right|^2 \label{part2}
\end{equation}
Remarkably the partitions functions (\ref{part1}) and (\ref{part2}) turn out
to be identical (we checked this fact numerically but it seems that the
analytical proof is a generalization of the Riemann's theta relations).
Again, we would like to stress that the identity $Z=Z'$ is
independent of $N$.

Now, let us concentrate on Green functions of both theories. It is
interesting to consider Green functions of local operators in the
noncommutative torus and $\theta$ dependent. Thus, consider the
following mass operators
\begin{equation}
\hat S_1(x)={\bar \psi}_R(x) \ast \psi_L(x) \; , \;\;\; \hat
S_2(x)={\bar \psi}_L(x) \ast \psi_R(x) \label{ops}
\end{equation}
and the v.e.v. of their product can be writen in the form
%
%
%
\begin{eqnarray}
\langle \hat S_1(\vec x)\; \hat S_2(\vec y) \rangle &=& \exp
\left(\frac{i\theta}{2}\; (\partial_{x_1}\partial_{x'_2} -
\partial_{x_2}\partial_{x'_1} + \partial_{y_1}\partial_{y'_2} -
\partial_{y_2}\partial_{y'_1})\right) \nonumber\\
&& \left. \langle \psi_L(\vec x') {\bar \psi}_L(\vec y)  \rangle
\right|_{\vec x'=\vec x} \;\; \left. \langle \psi_R(\vec y') {\bar
\psi}_R(\vec x)\rangle \right|_{\vec y' = \vec y}
\end{eqnarray}
The two-point function is the standard one for fermions on torus
of size $R$ with structure spin $\vec\alpha$
\begin{eqnarray}
\langle \psi_L(\vec x) {\bar \psi}_L(\vec y)  \rangle &=& \sum_{\vec k\in
\mathbb{Z}^2}\; \frac{\exp \left(2 \pi i (\vec k +\vec
\alpha)\cdot (\vec x - \vec y)/R\right) }{i ( k_z+\alpha_z)/R} \nonumber\\
\langle \psi_R(\vec x) {\bar \psi}_R(\vec y)  \rangle &=&- \sum_{\vec k\in
\mathbb{Z}^2}\; \frac{\exp \left(2 \pi i (\vec k +\vec \alpha)\cdot (\vec x -
\vec y)/R\right) }{i ( k_{\bar z}+\alpha_{\bar z})/R}
\label{tpf1}
\end{eqnarray}
where $k_z=k_1-i k_2$ and $k_{\bar z}=k_1+i k_2$. Finally, after a
straightforward computation we get
\begin{equation}
\langle \hat S_1(\vec x)\; \hat S_2(\vec y) \rangle =
-\sum_{\vec k, \vec k'} e^{\frac{4\pi^2\theta i}{R^2}\;(\vec k +
\vec \alpha)\wedge (\vec k' + \vec \alpha)} \; \frac{e^{2 \pi i
(\vec k -\vec k')\cdot (\vec x - \vec y)/R} }{( k_z+\alpha_z)(
k'_{\bar z}+\alpha_{\bar z})/R^2} \label{vev1}
\end{equation}
Now let us compute the corresponding correlator in the dual torus
$T'^2_{\theta=0}$.
Using (\ref{map1})-(\ref{map3}), the operators (\ref{ops}) are
mapped to
\begin{eqnarray}
S_1 &=& \sum_{\vec n, \vec m}^{N'-1} \;{\bar
\chi^{(\vec{n})}_R}(x)\;\chi^{(\vec{m})}_L(x) \;
J_{-(2\vec{n}+\vec{\alpha})} \cdot J_{2(\vec{m}+\vec{\alpha})}
\nonumber\\
S_2 &=& \sum_{\vec n, \vec m}^{N'-1} \;{\bar
\chi^{(\vec{n})}_L}(x)\;\chi^{(\vec{m})}_R(x) \;
J_{-2(\vec{n}+\vec{\alpha})} \cdot J_{2(\vec{m}+\vec{\alpha})}
\end{eqnarray}
thus the v.e.v. is mapped to
\begin{eqnarray}
\langle S_1(x) \otimes S_2(y) \rangle &=& \sum_{\vec n, \vec
n'}^{N'-1}\; \exp\left(\frac{2\pi i
M}{2N'}\,(2\vec n + 2\vec \alpha)\wedge (2\vec n' + 2\vec \alpha)\right)\times \nonumber\\
%
%
%
&& \hspace{-1cm}J_{2(\vec n- \vec n')} \otimes J_{2(\vec n'- \vec n)}
\langle \chi^{(\vec n)}_L(\vec x) {\bar \chi}^{(\vec n)}_L(\vec y)
\rangle \; \langle \chi^{(\vec n')}_R(\vec y') {\bar \chi}^{(\vec
n')}_R(\vec x)\rangle \label{vev2}
\end{eqnarray}
The two point functions appearing in the r.h.s of (\ref{vev2})  are the same
as that of equation (\ref{tpf1}) with a torus size $R/N'$ and spin structures
given by (\ref{spinsts}). Finally, substituting the two point functions into
equation (\ref{vev2}) and after a short computation we get
%
%
%
%
\begin{eqnarray}
\langle S_1(x) \otimes S_2(y) \rangle &=& -\sum_{\vec k, \vec k'}
\exp\left(\frac{4\pi^2\theta i}{R^2}\;(\vec k + \vec \alpha)\wedge
(\vec k' + \vec \alpha)\right)\times\nonumber\\
&& \hspace{-1cm}  \; \frac{\exp \left(2 \pi i (\vec k
-\vec k')\cdot (\vec x - \vec y)/R\right)}{( k_z+\alpha_z)(
k'_{\bar z}+\alpha_{\bar
z})/R^2} \; J_{2(\vec k- \vec k')} \otimes J_{2(\vec k'- \vec k)}
\label{vev3}
\end{eqnarray}
This  equation is identical to its noncommutative counterpart (\ref{vev1})
after the substitution (\ref{morita}), showing the Morita equivalence of
vacuum expectations values between dual theories. The generalization of this
result to more complex correlation functions is straightforward. But once we
have shown the equivalence of correlators in free theory, we can extend the
results, perturbatively, to any ``local" interaction of the fields. In fact,
it is clear that the structure and singularities of dual correlators are
identical, so the perturbative result to any order in dual theories will
coincide.

\section{Chiral anomaly and fermion determinant}

So far we have studied the Morita equivalence between free fermion theories
whose actions, being quadratic in the fields, are independent of $\theta$.
Even though this equivalence can be extended to arbitrary interactions in
perturbation theory, one can still think that there is some degree of
triviality in the examples arguing that we are just analyzing equivalences
between theories defined around free actions. However the
$\theta$-independence manifests only at the level of the actions; the
``local'' operators of the theory, as $\hat S_1$ and $\hat S_2$ in
(\ref{ops}), are of course $\theta$ dependent. In fact, for a fermion theory
in noncommutative space, the ``space of local interactions'' in the sense of
Wilson consists of all ``local'' star-product functionals of the fields.

Now let us study a less trivial example of the Morita equivalence, the chiral
anomaly and its relation to bosonization. In particular, we are going to show
the Morita equivalence between gauge effective actions of fermions coupled to
a gauge field, which is clearly a non perturbative result.

Consider a theory of fermions in the noncommutative torus
$T^2_{\theta}$ coupled minimally to a gauge field $A_{\mu}$. We
will work in the fundamental representation but other cases
(anti-fundamental or adjoint representation) are completely
analogous. The action is
\begin{equation}
S=-\frac{1}{8\pi}\int_{T^2_{\theta}}\,  d^2x\; \bar{\psi}\ast ( \;
\not\!\partial + \not\!\! A \ast)\; \psi + S[A_{\mu}] \label{action-g}
\end{equation}
where $S[A_{\mu}]$ is the gauge field action.

The gauge field satisfy periodic boundary conditions and can be
expanded  in Fourier modes defined in (\ref{modos}) as
\begin{equation}
A_{\mu}=\sum_{\vec k} A_{\mu}^{\vec{k}} U_{\vec{k}}
\end{equation}
%
%

The action (\ref{action-g}) is invariant under chiral
transformations \cite{MS1}, \cite{MS2}
\begin{equation}\label{chi}
\psi'(x) = {\cal U}_5(x)* \psi = \exp_*( \gamma_5 \alpha(x))* \psi
\end{equation}
%
%
%
and leads to the anomalous conservation of the chiral current
\begin{equation}
\partial_{\mu} j_{\mu}^5 = {\cal A} \; , \;\;\;\;\; j_{\mu}^5=
\psi^T * (\gamma^5 \gamma_{\mu})^T {\bar \psi}^T
\end{equation}
The chiral anomaly ${\cal A}$ can be calculated from the Fujikawa
Jacobian $J[\alpha]$ associated with an infinitesimal chiral
transformation ${\cal U} = 1 + \gamma_5 \delta \alpha$
\begin{equation}
\log J[\alpha]= -2  \int_{T^2_{\theta}} d^2x\;{\cal A}\; *
\delta\alpha\;\;, \;\;\;\;\;\;\; {\cal A} = \left.{\rm tr} \;
\gamma_5 \; \delta(x-x) \right\vert_{reg} \label{reg}
\end{equation}
To compute the infinitesimal Jacobian we parametrize the gauge
potentials with periodic fields $\phi$ and $\eta$ as
\begin{equation}
 \not\!\!A = \frac{i}{e}{\not\!\partial}{\cal U}[\phi,\eta]*
  {\cal U}^{-1}[\phi,\eta]+{\not\!\!A}^0 \label{par1}
\end{equation}
with
\begin{equation}
{\cal U}[\phi,\eta]=\exp_*(\gamma_5\phi+i\eta)
\end{equation}
and $A_\mu^0$ constants giving the Wilson phases around the cycles
of the torus.
We perform a change of the fermion variables
\begin{equation}
\psi = {\cal U}_t * \chi_t \;\; , \;\;\;\;\;\; \bar \psi = \bar
\chi_t
* {\cal U}_t^\dagger \; \; , \;\;\;\;\;
{\cal U}_t = \exp_*\left(t\left( \gamma_5\phi + i \eta \right)\right)\;,
%
\label{chat}
\end{equation}
and $t$ is a real parameter, $0 \leq t \leq 1$.  In particular, if
$t=1$ the fermions decouple from the gauge fields up to the
constant term $A_{\mu}^0$. This change of variables has associated
a Fujikawa jacobian $J[\phi,\eta;t]$ through the relation
\begin{equation}
\det{}_*\left(i\!\!\not\!\partial +e\!\not\!\!A\right)=
J[\phi,\eta;t]\det{}_* \left(i\not\!\!D[t]\right) \label{jac1}
\end{equation}
where
\begin{equation}
\label{Dt} \not\!\!D [t]=\not\!\partial + \not\!\partial{\cal
U}_{1-t}* {\cal U}_{1-t}^{-1} -i e {\not\!\!A}^0
\end{equation}
Differentiating and integrating over $t$ we finally have
\begin{equation}
\det{}_*\left(i\!\!\not\!\partial +e\!\not\!\!A\right)=
\det{}_*\left(i\!\!\not\!\partial +e\!\not\!\!A^0\right) \exp\left(-2\int_0^1
dt \int_{{T^2_{\theta}}}d^2x \,{\mathcal A}(t)*\phi\right) \label{jac3}
\end{equation}
where we have identified
\begin{equation}
-2\int_{{T^2_{\theta}}}d^2x \,{\mathcal A}(t)*\phi =
\frac{d}{dt}\left(\log_* J[\phi,\eta;t]\right) \label{ano}
\end{equation}
Equations (\ref{jac1}) and (\ref{jac3}) relate the Fujikawa
jacobian with the determinant of the Dirac operator and the chiral
anomaly.

Now, expression (\ref{reg}) has to be regularized in a
gauge-invariant way, thus as usual we write \cite{MS1},\cite{MS2}
\begin{equation}
{\mathcal A}(t)_{reg}=\lim_{M\to\infty}{\rm tr} \;\;\gamma_5
\frac{1}{R^2}\sum_{\vec k}{U_{\vec{k}+\vec{\alpha}}}^\dagger
 \exp_*\left({\not\!\!D[t]*\not\!\!D[t]}/{M^2}
 \right)U_{\vec{k}+\vec{\alpha}}
\end{equation}
and, after a straightforward calculation we obtain the usual result for the
chiral anomaly
\begin{equation}
{\mathcal A}(t)_{reg}= \frac{e}{2\pi}\,\varepsilon^{\mu\nu}
F^t_{\mu\nu}
\end{equation}
where $F^t_{\mu\nu}$ is the electromagnetic field strength tensor
\begin{equation}
F^t_{\mu \nu} = \partial_{\mu} A_{\nu}^t - \partial_{\nu} A_{\mu}^t -
ie(A_{\mu}^t * A_{\nu}^t - A_{\mu}^t * A_{\nu}^t)
\end{equation}
and $A_{\mu}$ is obtained from
\begin{equation}
\label{amut}
 \not\!\!A^t = \frac{i}{e}\not\!\partial {\cal
U}_{1-t}[\phi,\eta]* {\cal U}_{1-t}^{-1}[\phi,\eta]
\end{equation}
Finally, the fermion determinant takes the form
\begin{equation}
\det{}_*\left(i\!\!\not\!\partial +e\!\not\!\!A\right)=
\det{}_*\left(i\!\!\not\!\partial +e\!\not\!\!A^0\right) \;
e^{-\frac{e}{2\pi}\int_{{T^2_{\theta}}} d^2x\int_0^1 dt\,
\varepsilon^{\mu\nu} F^t_{\mu\nu}*\phi} \label{detcn}
\end{equation}
To compute the first factor of the above expression we note the
following. The constant field $A_{\mu}^0$ can be written as
\begin{equation}
A_\mu^0 = \frac{i}{e}U_{\vec \beta} * \partial_\mu (U_{\vec
\beta})^{-1}
\end{equation}
with $U_{\beta}$ defined as in (\ref{modos}) and $\vec \beta =
(R/2 \pi)\; {\vec A}_0$.
Then, performing a gauge transformation
%
$\psi \to \psi'=U_{\vec{\beta}} * \psi$,
%
we can eliminate the constant gauge field from the determinant at
the expense of changing the spin structures of the fermions $\vec
\alpha \to \vec \alpha + \vec \beta$.
Hence the determinant of the operator $\;i\!\!\not\!\partial +e\!\not\!\!A^0
\; $ is nothing but the partition function of a free fermion with spin
structure $\vec \alpha + \vec \beta$
\begin{equation}
\det{}_*\left(i\!\!\not\!\partial +e\!\not\!\!A^0\right) =
\frac{\left|\vartheta\left[^{\alpha_1 +\beta_1 -1/2}_{\alpha_2 +\beta_2
-1/2}\right] (0,\tau)\right|^2}{\left|\eta(\tau)\right|^2} \label{deta-1}
\end{equation}

\bigskip

At this point, we can compare these results with the corresponding
for the Morita-equivalent theory in the dual torus ${T^2}'$. The
action (\ref{action-g}) is mapped to the one
\begin{equation}
S=-\frac{N'}{16\pi}{\rm tr}_G \int_{{T^2}'} \,  d^2x\; \bar{\psi} ( \;
\not\!\partial + \not\!\! A)\; \psi + S[A_{\mu}] \label{action-2}
\end{equation}
where the fields $\psi$ are $N\times N$ fermion-valued matrices
and satisfy the boundary conditions (\ref{bc}). The $U(N)$ gauge
field $A_{\mu}$ is defines on the dual torus $T'^2_{0}$ and is
given by \cite{saraikin}-\cite{gura}.
\begin{equation}
A_{\mu}= \sum_{\vec n =0}^{N'-1}\;J_{\vec{n}} \sum_{\vec q\in \mathbb{Z}^2}
\; \exp\left(2\pi i N'\vec{q}\cdot\vec{x}/R\right) A_{\mu}^{\vec{q},\vec{n}}
U_{\vec n} \label{mapg}
\end{equation}
It satisfy the boundary conditions
\begin{equation}
A_{\mu}(x_1+R',x_2)=\Omega_1^\dagger A_{\mu}(x_1,x_2)\Omega_1 \; ,\;\;\;
A_{\mu}(x_1,x_2+R')=\Omega_2^\dagger A_{\mu}(x_1,x_2)\Omega_2
\label{bcg}
\end{equation}
with $\Omega_1$ and $\Omega_2$ defined in equation (\ref{omega}). Notice that
(\ref{bcg}) are constant gauge transformations so the action is
insensitive to them.

Using the mapped expressions for fermion fields, gauge fields and the
$t$-dependent transformation, we can deal with this fermion determinant
analogously. Indeed by a similar computation to the one used to obtain
(\ref{jac3}) we have
\begin{equation}
\det\left(i\!\!\not\!\partial +e\!\not\!\!A\right)=
\det\left(i\!\!\not\!\partial +e\!\not\!\!A^0\right)
\exp\left(-2\,\frac{1}{N}{\rm tr}_G \int_0^1 dt \int_{{T^2}'} d^2x
\,{\mathcal A}(t)\phi\right) \label{dets-na}
\end{equation}
where 
\begin{eqnarray}
{\mathcal A}(t)_{reg}&=&\lim_{M\to\infty}{\rm tr}
\left(\gamma_5\frac{1}{{R'}^2}\sum_{\vec q}\sum_{\vec n=0}^{N'-1}
\exp\left(-2\pi i(N'\vec q+\vec n+\vec \alpha)\vec x/R\right)
\; J_{2\vec n+2\vec \alpha}^\dagger\right. \times\nonumber \\
&&\left.\hspace{1.2cm} \exp\left(\frac{\not\!\!D[t]^2}{M^2}
\right)\exp\left(2\pi i(N'\vec q+\vec n+\vec \alpha)\vec
x/R\right) J_{2\vec n+2\vec \alpha}\right)
\end{eqnarray}
and $\phi$ can be written in terms of $A_{\mu}$ using equation
(\ref{amut}). Finally, after a standard computation we get the
well-known result for the nonabelian chiral anomaly in two
dimensions, which inserted in equation (\ref{dets-na}) gives
\begin{equation}
\label{detc} \det\left(i\!\!\not\!\partial +e\!\not\!\!A\right)=
\det\left(i\!\!\not\!\partial +e\!\not\!\!A^0\right)
\exp\left(-\frac{eN'}{4\pi}{\rm tr}_G\int_{{T^2}'} d^2x\int_0^1
dt\, \varepsilon^{\mu\nu} F^t_{\mu\nu}\phi\right)
\label{dets-na2}
\end{equation}
The first factor can be written analogously to the free fermion case as a
product of $N'^2$ partition functions of free fermions with spin structures
shifted by an amount $\vec \beta$, giving
\begin{equation}
\det\left(i\!\!\not\!\partial +e\!\not\!\!A^0\right) =
\prod_{n_1,n_2=0}^{N'-1}\left|\frac{\vartheta \left[^{(\alpha_1 +
\beta_1 + n_1)/N'-1/2}_{(\alpha_2 + \beta_2+n_2)/N'-1/2}\right]
(0,\tau)}{\eta(\tau)}\right|^2 \label{deta-2}
\end{equation}
As in the case of free fermion partition functions, it can be
checked numerically that  expressions (\ref{deta-1}) and
(\ref{deta-2}) are identical.

Using relation (\ref{integral}) and the Morita mapping for the gauge fields
it is immediate to show that the second factor in equation (\ref{dets-na2})
is exactly the same as its noncommutative counterpart in equation
(\ref{detcn}). Thus, both determinants are equal.

We can go farther and perform the integration of the $t$-parameter in the
anomaly equation (\ref{detcn}). In fact, this integration  can be done
without difficult in the light-cone gauge
%
%
\cite{MS1}-\cite{MS2}
\begin{eqnarray}
&&-\frac{e}{2\pi}\int_{T^2_\theta} d^2x\int_0^1 dt\, \varepsilon^{\mu\nu}
F^t_{\mu\nu}*\phi
%
= -\frac{1}{8\pi} \int_{T^2_\theta} d^2x \left(\partial_\mu g^{-1} \right)
* \left(\partial_\mu g\right)
\nonumber\\
%
%
&& \hspace{1.0 cm}+ \frac{i}{4\pi} \epsilon_{ij}\int_{T^2_\theta}
d^2x\int_0^1 dt\; g^{-1}
* \left(\partial_i g\right)
* g^{-1} * \left(\partial_j g\right) * g^{-1} *
\left(\partial_t g \right) \label{WZ}
\end{eqnarray}
which is the Moyal deformation of the Wess-Zumino-Witten (WZW) action. This
action is highly nonlinear and non-perturbative in nature. It is well known
that, in ordinary space, the correlators of the WZW can be computed exactly,
by invoking the infinite dimensional symmetries of the theory, namely the
Virasoro algebra and the affine current algebra. However, it is not known how
to solve this problem in noncommutative space as the star deformations of the
Virasoro and affine algebras are not fully understood. Some progress in this
direction was done in \cite{MS2} where it was shown that, through a
Seiberg-Witten mapping, the  noncommutative WZW action is mapped to an
ordinary space, $U(1)$ WZW model (a $U(1)$ WZW model in ordinary space is
equivalent to a free masless boson theory). Notice nevertheless that the
Seiberg-Witten mapping is not an isomorphism.

But in the noncommutative torus the situation is different; the Morita
equivalence give us an isomorphism between noncommutative algebras and in the
special case of a rational $\theta$ parameter, one of the isomorphic theories
is a commutative one. Since we know that the gauge effective action of
noncommutative theory of masless Dirac fermions coupled to a gauge field is a
star deformed WZW theory and we have shown that, through the Morita
equivalence, is mapped to an ordinary space WZW theory, we have an actual
isomorphism between both theories. That is, we can give meaning to the
concept of a ``noncommutative conformal field theory", as the Morita
equivalent version of an ordinary-space CFT.

Finally, it is not difficult to show \cite{MS1}, \cite{nunez},
\cite{MS2} from eqs. (\ref{dets-na2}),(\ref{WZ}), that the
bosonization of a free fermion theory is a star deformed WZW
theory. It would be interesting to perform the Morita mapping to
this theory and compare it with the bosonization of the ordinary
space non-abelian fermion theory. We hope to report on this issue
in a future work.

\vspace{1 cm}

\noindent\underline{Acknowledgements}:  This work was partially supported  by
UNLP, CONICET (PIP 4330/96), ANPCYT (PICT 03-05179), Argentina. E.F.M. was
partially supported by Fundaci\'on Antorchas, Argentina.


\newpage


\end{document}